\documentclass[aip,jcp,superscriptaddress,twocolumn]{revtex4}

\usepackage{graphicx}       % Include figure files
\usepackage{dcolumn}        % Align table columns on decimal point
\usepackage{bbm}
\usepackage{amsmath}        
\usepackage{amssymb}
\usepackage{epsf}
\usepackage{subfigure}
\usepackage{graphics}
\usepackage{rotating}
\usepackage{bm}

\begin{document}

\title{Free energy calculations from adaptive molecular dynamics simulations with adiabatic reweighting}

\author {Lingling Cao}
\affiliation{CEA, DEN, Service de Recherches de M\'etallurgie Physique, F-91191 Gif-sur-Yvette, France}

\author{Gabriel Stoltz}
\affiliation{Universit\'e Paris-Est, CERMICS (ENPC), INRIA, F-77455 Marne-la-Vall\'ee, France}

\author{Tony Leli\`evre}
\affiliation{Universit\'e Paris-Est, CERMICS (ENPC), INRIA, F-77455 Marne-la-Vall\'ee, France}

\author{Mihai-Cosmin Marinica}
\affiliation{CEA, DEN, Service de Recherches de M\'etallurgie Physique, F-91191 Gif-sur-Yvette, France}

\author{Manuel Ath\`enes}
\affiliation{CEA, DEN, Service de Recherches de M\'etallurgie Physique, F-91191 Gif-sur-Yvette, France}

\begin{abstract}
We propose an adiabatic reweighting algorithm for computing the free energy along an external parameter from adaptive molecular dynamics simulations. The adaptive bias is estimated using Bayes identity and information from all the sampled configurations. We apply the algorithm to a structural transition in a cluster and to the migration of a crystalline defect along a reaction coordinate. Compared to standard adaptive molecular dynamics, we observe an acceleration of convergence. With the aid of the algorithm, it is also possible to iteratively construct the free energy along the reaction coordinate without having to differentiate the gradient of the reaction coordinate or any biasing potential. 
\end{abstract}

\maketitle

An important task of molecular simulation in material science, chemistry or biophysics is the computation of the free energy $\mathcal{A}(\zeta)$ along an external parameter $\zeta$ that may be inverse temperature, pressure or a chemical potential. In these situations, the free energy is useful for characterizing the conditions of phase equilibria, for example between the solid-like and liquid-like states of an atomic cluster. A related quantity that is frequently desired is the free energy along a reaction coordinate $\xi(q)$, where $\xi$ is a function of the internal degrees of freedom, here the position $q$ of the system. In systems presenting broken ergodicity or metastabilities resulting from rare crossings of entropic or energetic barriers, the free energy along $\xi(q)$ is often used within transition state theory~\cite{chandler:1987} to estimate the rates of barrier crossings. Those may be the jump frequencies of a defect in a crystal to give a second practical example. 

Basic techniques allowing to compute $\mathcal{A}(\zeta)$ from Monte Carlo or molecular dynamics (MD) simulations~\cite{frenkel2002understanding,wales:2003} are thermodynamic integration, free energy perturbation and nonequilibrium work methods.~\cite{jarzynski:1997,jarzynski:2007} To improve the accuracy of the results, a reweighting procedure~\cite{bennett:1976,ferrenberg:1989,tan:2004,shirts:2008,habeck:2012} is often implemented to post-process and combine the data harvested in multiple simulations performed with different values of the external parameter. Reweighting applies in particular to rare barrier crossing problems, in which case external parameters are introduced to restrain the system across the barrier via harmonic coupling to the reaction coordinate.~\cite{torrie:1977,kastner:2011,lelievre:2010} It then allows to obtain the free energy of the reaction coordinate without having to evaluate its second-order derivatives. These second-order derivatives are often difficult to compute and appear when the free energy is differentiated with respect to the reaction coordinate, as in the constrained thermodynamic integration~\cite{carter:1989} method or in the adiabatic free energy dynamics~\cite{rosso:2002} (AFED) method. 

A common feature of reweighting algorithms is that, in order to minimize the statistical variance, the information of any configuration sampled at a given value of the external parameter is included in the estimators for the free energies at all values of the external parameter. However, this information is not used in the course of the simulations to improve the construction of the samples. In this communication, we show how to adaptively perform the sampling by reweighting the information contained in all configurations previously generated. We focus here on the adaptive biasing force (ABF) framework.~\cite{darve:2001,darve:2008}

In ABF methods,~\cite{darve:2001,darve:2008} a biasing force is adapted and used in the molecular dynamics to achieve uniform sampling of the chosen reaction coordinate. We thus consider that $\zeta$ is a reaction coordinate taking values in $\mathcal{Z}$ and write $\mathcal{A}$ as a potential of mean force:~\cite{lyubartsev:1992} 
\begin{equation} \label{eq:free_energy}
\mathcal{A} (\zeta_\star) = - \beta ^{-1} \ln \left( \int_{\mathcal{Z} \cup \mathcal{Q}} \mathbbm{1}({\zeta_\star} | \zeta) e^{ - \beta U (\zeta,q)} d\zeta dq \right)
\end{equation}
where $\beta^{-1}$ denotes the reference temperature, $q$ the coordinates of the multi-particle system, $\mathcal{Q} $ its phase space, $\mathbbm{1}({\zeta_\star}| \cdot)$ the characteristic function of the histogram bin containing $\zeta_\star$ and $U(\zeta,q)$ the extended potential. This one usually takes the form $U(\zeta,q) = V(q)+ \zeta E(q)$ in alchemical free energy calculations, where $V$ and $V+E$ are the reference and target potentials, respectively.  In another common simulation set-up, $\zeta$ is a restraining parameter harmonically coupling to a reaction coordinate $\xi(q)$ via $U(\zeta,q) = V(q)+ \frac{1}{\eta} |\zeta - \xi(q)|^2$ where $\eta^{-1}$ is a spring stiffness. Restraining potentials with functional forms different from linear or quadratic in $\zeta$ may also be used. Let $\partial_\zeta U$ and $\nabla_q U$ denote the derivatives of $U$. Since $\mathcal{Z}$ is independent of $\mathcal{Q}$, the mean force $\mathcal{A}^\prime(\zeta)$ is formally equal to 
\[
\mathbb{E}\left[\partial_\zeta U\left( \zeta , \cdot\right) | \zeta \right] = \frac{\int_\mathcal{Q} \partial_\zeta U( \zeta ,q) e^{ - \beta U (\zeta,q)} dq }{\int_\mathcal{Q} e^{ - \beta U (\zeta,q)} dq}, 
\]
the conditional expectation of $\partial_\zeta U$ given $\zeta$. 
The biasing force used in ABF at time $t$ is $A^\prime_t$, the current estimate of $\mathcal{A}^\prime$ obtained here using the histograms accumulated along the past trajectory. Denoting the system coordinates at time $t$ by $(\zeta_t, q_t)$, an ABF algorithm in the extended system (ABF-E) writes~\cite{lelievre:2010}
\begin{subequations} \label{eq:dynABF_standard}
\begin{align} 
A_t^\prime(\zeta_t) & = \frac{\int_0^t \partial_\zeta U(\zeta_s,q_s) \mathbbm{1}({\zeta_t} | \zeta_s) ds }{\tau + \int_0^t \mathbbm{1} (\zeta_t | \zeta_s )  ds }, \label{eq:running_average} \\
\label{eq:dyn_zeta} d\zeta_t   & =  \left[ A_t^\prime(\zeta_t) -  \partial_\zeta U \left( \zeta_t, q_t \right) \right] dt + \sqrt{2 / \beta } d\bar{W}_t, \\
\label{eq:dyn_q} dq_t       & = - \nabla_q U \left[ \zeta_t, q_t \right] dt + \sqrt{2/\beta } dW_t . 
\end{align}
\end{subequations}
The bias $A^\prime_t$ is set to $0$ outside $[\zeta^{\mathrm{min}},\zeta^{\mathrm{max}}]$ interval. The Langevin dynamics~(\ref{eq:dyn_zeta},\ref{eq:dyn_q}) is driven by Wiener process $(d\bar{W}_t,dW_t)$ and by a biasing force converging to $\mathcal{A}^\prime$ within $[\zeta^{\mathrm{min}}, \zeta^{\mathrm{max}}]$ in the long time limit.~\cite{lelievre:2008} The time $\tau$ in the denominator is a small factor (that may be zero or decay to zero~\cite{lelievre:2008b,henin:2010}) introduced to prevent an initially too large bias from driving the dynamics out of equilibrium. The dynamics~(\ref{eq:dynABF_standard}) is such that, at convergence, the variable~$\zeta_t$ freely explore the interval~$[\zeta^{\mathrm{min}},\zeta^{\mathrm{max}}]$.

Here, we use Bayes formula to more efficiently estimate the running average~\eqref{eq:running_average} associated with the histogram bins of~$\zeta$. Let us first consider a time-independent biasing potential $A(\zeta)$. The biased potential is therefore $U_{A} (\zeta,q)= U(\zeta,q) -A (\zeta)-\varphi_A$, the constant $\varphi_A$ normalizing the canonical distribution at temperature $\beta^{-1}$. Thus, the joint probability of $(\zeta,q)$ is $e^{-\beta U_A(\zeta,q)}$. 
The marginal probability of $q$ is $\bar{\mathrm{P}}_{A} (q) = \int_\mathcal{Z} e^{-\beta U_A(\zeta,q)} d\zeta $ and the conditional probability of $\zeta$ given $q$ is $\bar{\pi}_A(\zeta | q)  = e^{-\beta U_A(\zeta,q)}/\bar{\mathrm{P}}_A(q) $. Denoting the conditional expectation associated with $\bar{\pi}_A$ by $\overline{\mathbb{E}}_A $, we have 
\begin{eqnarray}~\label{eq:field}
\overline{\mathbb{E}}_A \left[ \nabla_q U( \cdot , q ) | q \right] = \int_\mathcal{Z} \nabla_q U ( {\zeta } , q ) e^{-\beta U_A({\zeta} , q )} d\zeta / \bar{\mathrm{P}}_A (q) \nonumber , 
\end{eqnarray}
which corresponds to minus the gradient of $\beta^{-1} \ln \bar{\mathrm{P}}_A (q)$. Therefore, the following Langevin dynamics 
\begin{equation}
 dq_t =   - \overline{\mathbb{E}}_A \left[ \nabla_q U( \cdot , q_t ) | q_t \right] dt + \sqrt{2/\beta} dW_t \label{eq:dyn_homogene}
\end{equation}
samples $\bar{\mathrm{P}}_{A} (q)dq$. 
Even though $\zeta$ is not propagated in~\eqref{eq:dyn_homogene}, the conditional probability of $q$ given $\zeta$ may be evaluated from the conditional probability of $\zeta$ given $q$ using Bayes formula
\begin{equation}\label{eq:bayes}
\pi (q | \zeta )  = \frac{\bar{\pi}_A (\zeta | q ) \bar{\mathrm{P}}_{A} (q) } { \int_\mathcal{Q} \bar{\pi}_A(\zeta | \tilde{q}) \bar{\mathrm{P}}_A(\tilde{q}) d\tilde{q}}. 
\end{equation}
The integral in the denominator defines $\mathrm{P}_A(\zeta)$, the marginal probability of $\zeta$. 
Owing to the ergodic theorem and to Bayes formula~\eqref{eq:bayes}, the conditional expectation of any observable $\mathcal{O}(\zeta,q)$ given $\zeta$ can be estimated from    
\begin{equation}
\mathbb{E}\left[\mathcal{O}(\zeta,\cdot)|\zeta \right] = \lim_{t \rightarrow + \infty } \frac{ \int_0^t  \mathcal{O}(\zeta,q_s) \bar{\pi}_{A} (\zeta | q_s)ds} { \int_0^t \bar{\pi}_{A} (\zeta | q_s) ds } \label{eq:expect_homogene},
\end{equation}
where $\left\{ q_s \right\} _{0 \le s \le t}$ is a long trajectory generated using~\eqref{eq:dyn_homogene}. 
The scheme implementing the expectation form of Bayes indentity~\eqref{eq:expect_homogene} is called ``adiabatic reweighting''. Here,  adiabaticity refers to the virtual dynamical decoupling that is involved: when $q_s$ evolves very slowly compared to $\zeta_s$, the latter variable has enough time to fully explore its subspace $\mathcal{Z}$ and visits any value of $\zeta$ with the current equilibrium probability $\bar{\pi}_{A} (\zeta | q_s)$. Concommitantly, the force exerted upon $q_s$ is $- \overline{\mathbb{E}}_A \left[ \nabla_q U( \cdot , q_s ) | q_s \right]$, the average of $-\nabla_q U( \zeta , q_s )$ taken over the current equilibrium distribution of $\zeta$ given $q_s$. To perform ABF simulations with adiabatic reweighting, we first notice that setting $\mathcal{O}$ to  $\partial_\zeta U$ in~\eqref{eq:expect_homogene} yields an estimate of $\mathcal{A}^\prime$. We next, in analogy with~\eqref{eq:dynABF_standard}, suggest to adapt the biasing force $A^\prime_t$ by setting $\bar{\pi}_A$ to $\bar{\pi}_{A_s}$ in~\eqref{eq:expect_homogene} and $\overline{\mathbb{E}}_A$ to $\overline{\mathbb{E}}_{A_t}$ in~\eqref{eq:dyn_homogene}. This leads to a new ABF scheme (for $\zeta \in [\zeta^{\mathrm{min}}, \zeta^{\mathrm{max}} ]$) 
\begin{subequations} \label{eq:dynABF_BT}
\begin{align}
\label{eq:deno} A^\prime_t (\zeta)  & = \frac{\int_0^t \partial_\zeta U\left(\zeta, q_s \right) \bar{\pi}_{ A_s} (\zeta | q_s) ds }{\tau + \int_0^t \bar{\pi}_{ A_s} (\zeta | q_s )  ds }, \\
 \label{eq:dyn_inhomogene} dq_t            & = - \overline{\mathbb{E}}_{A_t}\left[ \nabla_q U(\cdot,q_t)| q_t \right] dt + \sqrt{2/\beta} dW_t.  
\end{align}
\end{subequations}
The resemblance with the original ABF-E scheme is striking, except that $\zeta$ is not dynamically propagated. 
The computation of $A^\prime_t$ and its numerical integration to obtain $A_t$ are performed at each time step on the same grid. 
$N_\zeta=10^3$ grid points are used. 
From ${A_t}$, the weights $\pi_{A_t}$ are evaluated and used to average $\nabla_q U (\cdot,q_t )$ in~\eqref{eq:dyn_inhomogene} (on the grid) and to update the numerator and denominator in~\eqref{eq:deno} for the next step. The computational cost of these operations is typically a small fraction of the one required for evaluating and differentiating the potential energies. 

Note that, as an alternative to the ABF framework, we may adapt the biasing potential~\cite{brukhno:1996,wang:2001,laio:2002,marsili:2006} rather than its gradient. An adaptive biasing potential method with adiabatic reweighting consists in replacing~\eqref{eq:deno} by 
\begin{equation}
 \frac{d A_t (\zeta)}{dt} = - \omega \beta^{-1} \ln \left[1 +  \tau ^{-1} \int_0^t \bar{\pi}_{ A_s} (\zeta | q_s )  ds \right]
\end{equation}
where the positive constant $\omega$ is the updating rate. Tuning this additional simulation parameter is not straightforward. Hence, we restrict the present investigation to the ABF framework. Note also that the AFED method enforces an adiabatic decoupling that is opposite to the one involved in Eqs.~\eqref{eq:expect_homogene} or~\eqref{eq:deno}: the particle system is therein assumed to fully explore the conditional distribution of $q$ given the instantaneous value of the reaction coordinate, whose dynamics is, owing to a large damping coefficient, slowed down and decoupled from the ones of the remaining coordinates.~\cite{rosso:2002}

The ABF algorithm with adiabatic reweighting (ABF-AR) is now applied to two benchmark systems illustrating respectively the alchemical and reaction coordinate cases. The alchemical application aims at characterizing the thermodynamic structural transition between the liquid-like and solid-like states of LJ55, a cluster system consisting of 55 particles interacting via a Lennard-Jones potential.~\cite{wales:2003,lynden-bell:1994} The global energy minimum is $-279.248 \epsilon_{LJ}$ where $\epsilon_{LJ}$ is the depth of the Lennard-Jones potential well. Its struture is a Mackay icosahedron. The cluster potential energy is $E$ and the reference potential is 0, so that $\beta U(\zeta,q)=\zeta E(q)$. From $\beta \partial_\zeta U = E$, it follows that $\beta \mathcal{A}^\prime$ is the mean potential energy and that $\beta \zeta^{-1} \mathcal{A}(\zeta)$ is the Helmholtz free energy (up to an additive constant) at effective temperature $T=(\beta \zeta)^{-1}$. 
Let $S(E)$ denote the microcanonical entropy and $\mathrm{P}(E|\zeta) = \exp\left[ S(E) + \beta \mathcal{A}(\zeta)-\beta\zeta E \right]$ be the conditional probability of $E$ given $\zeta$. The signature of a thermodynamic transition is the presence of an inflection point on the curve $\left\{ \zeta, \mathcal{A}^\prime(\zeta) \right\}$, or of a loop on the (van der Waals) curve $\left\{ \beta^{-1} S^\prime(E), E \right\}$ where $S^\prime=dS/dE$. These two curves are the locus of the stationary points satisfying $\partial_\zeta \mathrm{P}(E|\zeta)=0$ and $\partial_E \mathrm{P}(E|\zeta) = 0$, respectively. 

\begin{figure}
\includegraphics[scale=0.7]{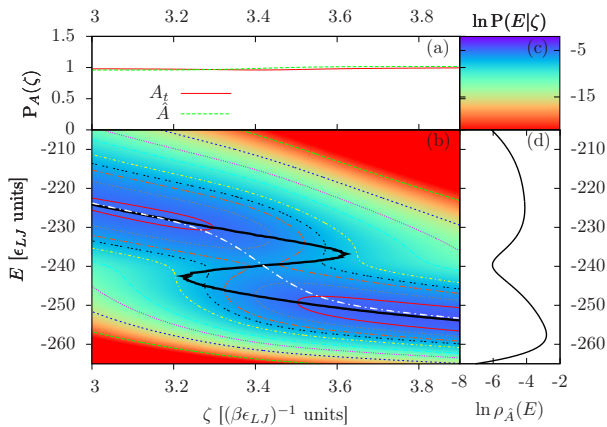}
 \caption{(a) $\mathrm{P}_A(\zeta)$ is scaled to the uniform distribution, as estimated in the adaptive run ($A_t$)  and production run ($\hat{{A}}$); (b) contour map of $\ln \mathrm{P}(E | \zeta)$ with some isolines, and with color-box displayed in (c); the thick white dashed and thick black solid lines represent the $\left\{ \zeta, \mathcal{A}^\prime(\zeta) \right\}$ and $\left\{ \beta^{-1} S^\prime(E), E \right\}$ curves, respectively; (d) $\left\{ \ln \rho_{\hat A }(E), E \right\}$ curve. Lennard-Jones units are used.} 
\label{fig:contour}
\end{figure}

First, we obtain an estimate of the free energy by generating an ABF-AR dynamics of $10^9$ steps using $\Delta t = 10^{-4}$ in Lennard-Jones units and $\tau=0$. We set $\beta^{-1}=0.3 \epsilon_{LJ}$ and $\mathcal{Z} = [2.5/(\beta \epsilon_{LJ} ) , 5/(\beta \epsilon_{LJ} )]$. The biasing force $A^\prime_t(\zeta)$ is estimated within $\mathcal{Z}$. The final biasing potential obtained by this procedure is denoted by $\hat{A}$. In a second step, we perform a production run of the same duration as before replacing $A$ by $\hat{A}$ in~\eqref{eq:dyn_homogene}, and estimate $\mathrm{P}(E|\zeta)$ by adiabatic reweighting~\eqref{eq:expect_homogene} where the observable is replaced by the characteristic function $\mathbbm{1}_{E,E+\Delta E}$. The histogram bin width is $\Delta E=0.1$. We also estimate $\mathrm{P}_A (\zeta)$ by averaging $\bar{\pi}_A(\zeta | q_t)$ in the adaptive run~\eqref{eq:dynABF_BT} and production run~\eqref{eq:dyn_homogene}. The density of states $g(E) = \exp [S(E)]$ is proportional to $\rho_{\hat{A}}(E) /\mathrm{P}_{\hat{A}} (q_E) $, the histogram of the sampled energies divided by the marginal probability of any configuration of energy $E$. This proportionality relation with the sampled data of the production run is used to construct the $\left\{\beta^{-1} S^\prime(E),E\right\}$ and $\left\{ \zeta, \mathcal{A}^\prime \right\}$ curves (standard reweighting~\cite{lee:1993}). 

Results are displayed in Fig.~\ref{fig:contour}. Panel (a) shows that the estimated marginal probability of $\zeta$ is flat over $\mathcal{Z}$: the converged biasing forces fully compensate the mean forces. Results for $\mathrm{P}(E|\zeta)$, up to a normalizing factor, are shown in the contour map of panel (b). The stationary points on the isolines are perfectly located on the superimposed $\left\{ \zeta, \mathcal{A}^\prime(\zeta) \right\}$ and $\left\{ \beta^{-1} S^\prime(E), E \right\}$ curves. Adiabatic and standard reweighting techniques yield matching results. The van der Waals loop clearly evidences the liquid-solid transition. 
The distribution of the sampled energies, shown in panel (c), is bimodal. This feature results from the phase coexistence occurring at intermediate inverse temperatures around $\zeta_c=3.42$. 

\begin{figure}
\includegraphics[scale=0.74]{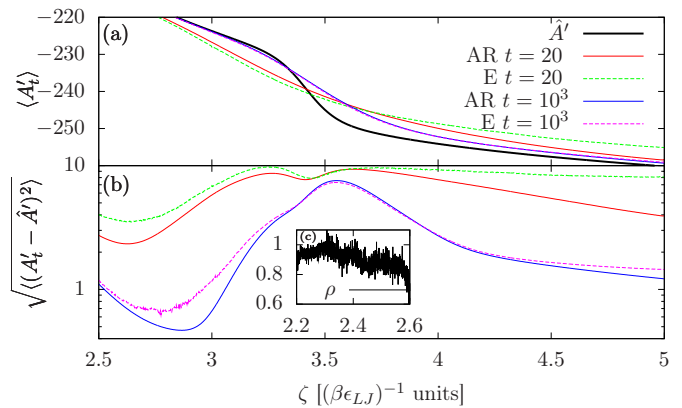}
\caption{ABF simulations with adiabatic reweighting (AR) or extended dynamics (E): (a) trajectory averages of the biasing force along $\zeta$ at two times $t$, compared to $\hat{A}^\prime$, the reference mean force; (b) error defined as a standard deviation from the reference curve $\mathcal{A}^\prime$; (c) $\rho$ represents the reduction of statistical variances resulting from adiabatic reweighting, estimated at $t=10^4$ from 200 trajectories. Energies are given in $\epsilon_{LJ}$ units. }
\label{fig:comparison}
\end{figure}

To assess the efficiency of ABF-AR, we make a numerical comparison with ABF-E. 
We generate $10^3$ dynamical trajectories of $10^7$ time-steps (of duration $10^3$) with both ABF-AR and ABF-E. Initial configurations are drawn from the canonical distribution at $\beta=0.3$. 
The trajectory average of $A^\prime_t$ and a measure of the average error are displayed in panels (a) and (b) of  Fig.~\ref{fig:comparison}, respectively, both as a function of $\zeta$ and at two times. We observe a faster convergence with adiabatic reweighting: the speed-up is significant at the early stage $t=20$ but more moderate at later times $t=10^3$. We estimate the asymptotic statistical variance of ABF-AR and ABF from 200 long runs of duration $t=10^4$ in a narrower temperature range. The reduction $\rho$ of the statistical variance displayed in panel (c) of Fig.~\ref{fig:contour} is around $15\%$. 

In the reaction coordinate application, $\xi(q)$ describes the migration of a vacancy in $\alpha$-Iron, a rare event on the femtosecond scale. The rate of vacancy migration is controlled by the free-energy barrier overcome by an atom jumping into a neighboring vacancy. The simulation set-up~\cite{athenes:2010,athenes:2012} is as follows: an embedded atom model potential~\cite{ackland2004} describes the atomic interactions $V$ and the reaction coordinate is the projection of the migrating atom into a $\langle 1 1 1 \rangle$ direction of the bcc crystalline structure that is aligned with the initial sites of the vacancy and of the jumping atom. The mean force along $\xi$ is a conditional expectation in $\mathcal{Q}$~\cite{darve:2001,sprik:1998,otter:1998,ciccotti:2008}:
\begin{equation} \label{eq:std_conditional}
 \mathcal{F}^\prime (\xi_\star) = \mathbb{E} \left[ \left. \frac{ \nabla V \cdot \nabla \xi }{|\nabla \xi |^{2}} - \frac{1}{\beta} \nabla \cdot \frac{\nabla \xi} { |\nabla \xi |^{2}} \right| \xi(q) = \xi_\star \right].
\end{equation}
This different notation is used to avoid confusion with $\mathcal{A}^\prime (\zeta)$. A reference free energy profile along $\xi(q)$ is computed at $T=500\ \mathrm{K}$ with ABF which uses $\nabla \xi(q_t) \cdot F^\prime_t (\xi(q_t))$ as biasing force, where $F^\prime_t$ is the current estimate of $\mathcal{F}^\prime$ at $\xi_\star = \xi(q_t)$. A first run of $2\cdot 10^{7}$ time-steps (of duration $t_1$) is generated. Then, $F_{t_1}^\prime$ is frozen and $10^2$ production runs of same duration are generated to construct $\mathrm{p}(\xi) = \langle \mathbbm{1}_\xi \rangle$, the occupation histograms of $\xi$. Our reference free energy is $\hat{F}(\xi)=F_{t_1}(\xi)-\beta^{-1}\ln \mathrm{p}(\xi)$ where $F_{t_1}$ is an integral of $F_{t_1}^\prime$. 

\begin{figure}
\includegraphics[scale=0.6]{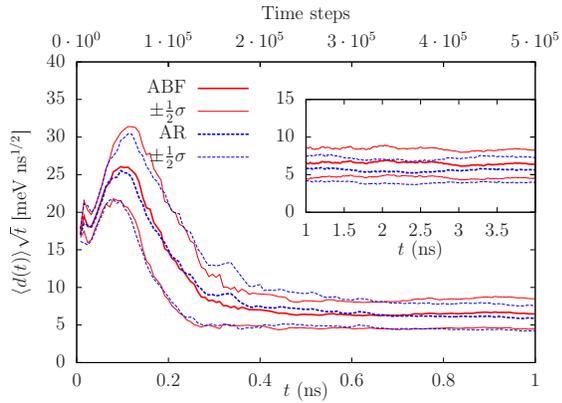}
\caption{Time evolution of $\langle d (t) \rangle \sqrt{t}$ for ABF and ABF-AR (AR); $\langle d(t) \rangle $ is the error averaged over 100 independent runs.} 
\label{fig:convergence}
\end{figure}

In ABF-AR simulations, $\zeta$ is an additional parameter controlling the volume visited by $\xi(q)$ via a harmonic coupling of stiffness $\eta^{-1}$. We use $\tau=\Delta t/N_\zeta$ in~\eqref{eq:dynABF_BT} where $\Delta t$ is the time-step. 
The derivatives of $U(\zeta,q)$ write  $\partial_\zeta U (\zeta,q) = \left[\zeta - \xi(q)\right]/\eta $ and $\nabla_q U = \nabla V - \nabla \xi \partial_\zeta U$.  We set $\eta=3.20 \cdot 10^{-2} \mathrm{\AA^2 / eV }$ in order to have a strong coupling between $\zeta$ and $\xi$ (compare $\sqrt{\eta T} = 3.71 \cdot 10^{-2} \textrm{\AA}$ and the cubic unit cell parameter $a_0=2.8553 \textrm{\AA}$). 
%Pulling on $\zeta$ thus enables the thermostated dynamics to cross the free energy barrier. 
The strong coupling does not affect the convergence of $F^\prime_t$ compared to ABF. To show this, we monitor $F^\prime_t$ using both ABF-AR and ABF, integrate to obtain $F_t$ and measure the distance to $\hat{{F}}$ using $d(t) = \frac{1}{n} {\sum }_{\ell=1}^{n} | F_t(\xi_\ell) - \hat{{F}}(\xi_\ell)|$
where the $\xi_\ell$ are the positions of the $n$ histogram bins. The distance is averaged over 100 runs. 
The error $\langle d (t) \rangle \sqrt{t}$ is plotted in Fig.~\ref{fig:convergence} as a function of time: both adaptive MD simulations exhibit the same transient regime and the same plateau value. 

\begin{figure}
\includegraphics[scale=0.8]{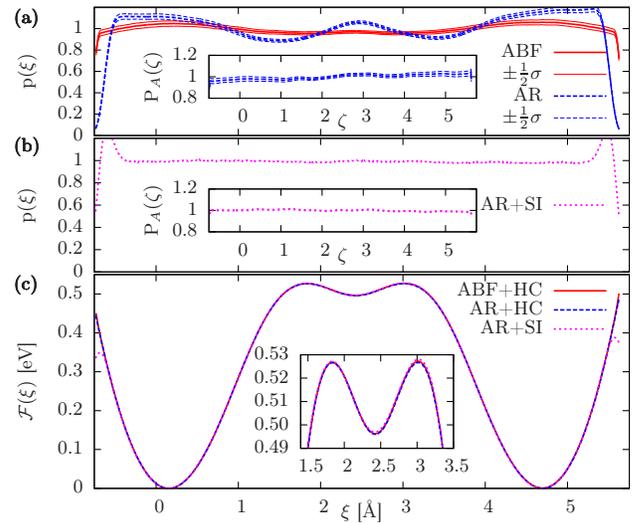}
\caption{(a) scaled histograms $p(\xi)$ of $\xi$ for ABF and ABF with adiabatic reweighting (AR). Error bars correspond to $\sigma$, the standard deviation of the 100 runs; $\zeta$-histogram in the insert is evaluated with AR and with frozen $A$ to $A_{t_1}$; (b) histogram of $\xi$ after a second iteration of ABF-AR (see text); inserts of panels (a) and (b) display the corresponding $\zeta$-histograms; (c) Estimated free energy with histogram correction (HC) or second iteration (SI), with a zoom on the barrier (insert).} 
\label{fig:comparison_RC}
\end{figure}

Estimating the mean force from Eq.~\eqref{eq:std_conditional} is often tedious due to the presence of the second-order space derivatives of $\xi(q)$. In practice, $\mathcal{F}^\prime(\xi)$ may also be estimated using Hamiltonian dynamics with only the first-order derivatives of $\xi(q)$ with respect to space and time.~\cite{darve:2008} Here, two additional alternatives to~\eqref{eq:std_conditional} are proposed for estimating $\mathcal{F}^\prime(\xi)$ using Langevin dynamics. The estimation can be achieved by histogram correction, noticing that the biasing potential $F_{t_1}\left[\xi(q)\right]$ in ABF is to be replaced in ABF-AR by the integrated bias $B\left[\xi(q)\right] = \beta^{-1} \ln \left[ \bar{\mathrm{P}}_{A_{t_1}}(q)/\bar{\mathrm{P}}_{0}(q)\right]$ where $\bar{\mathrm{P}}_{0}(q)$ denotes the unbiased probability of $q$. We thus construct the occupation histograms of both $\xi$ and $\zeta$ from $10^2$ production runs with frozen biasing forces $A^\prime_{t_1}$ and same duration. Results are displayed in Fig.~\ref{fig:comparison_RC}.a. Error lines are plotted from the estimated standard errors $\sigma$.
The histograms in $\xi$ for ABF and in $\zeta$ for ABF-AR are reasonably flat given the relatively short duration of the two simulations. The $\mathrm{p}(\xi)$ histogram for ABF-AR, shown in Fig.~\ref{fig:comparison_RC}~(a), is not flat and rather characterizes a small residual barrier resulting from the additional spring. The ABF-AR estimate of $\mathcal{F}(\xi)$ is $B_{t_1}(\xi)-\beta^{-1}\ln \mathrm{p}(\xi)$. 

Our second alternative to~\eqref{eq:std_conditional} consists in adapting the potential $U_A$ along $\xi$ using the integrated bias $B$. Let $\bar{\mathrm{P}}_A^B(q)$ denote the marginal probabability of $q$ associated with the updated potential $U_A-B\circ \xi$. We perform a second ABF-AR simulation of duration $t_2=10 \times t_1$ to obtain $A_{t_2}$ starting from $A_{t=0}=0$ with the updated potential. We observe that the histogram $\mathrm{p}(\xi)$, shown in Fig.~\ref{fig:comparison_RC}~(b), is flattened. 
As a result, the new bias, integrated from the relation $C \left[\xi(q)\right] = \beta^{-1} \ln[\bar{\mathrm{P}}^B_A(q)/\bar{\mathrm{P}}_0(q)]$
where $A = A_{t_2}$, is expected to be an improved estimate of the free energy $\mathcal{F}$ over the migration barrier.  
The excellent agreement between the $\mathcal{F}(\xi)$ estimates obtained after the second iteration and from histogram correction is shown in Fig.~\ref{fig:comparison_RC}~(c). With ABF-AR, the residual barrier after the first iteration being 1.44\% the estimated free energy barrier, the residual error after the second iteration is expected to be negligible compared to the statistical error of the sampling. 

In conclusion, we observe that adiabatic reweighting accelerates the initial convergence of the biasing forces along the external parameter in adaptive MD simulations. Moreover, with the aid of the reweighting algorithm, it is also possible to iteratively construct the free energy of a reaction coordinate without differentiating its gradient or any biasing potential. 
Whenever the reaction coordinate is not differentiable, the Langevin dynamics that was employed in the present study is to be replaced by a Metropolis algorithm. Adiabatic reweighting may also be combined with the waste-recycling Monte Carlo approach~\cite{frenkel:2004,frenkel:2006,athenes:2007,athenes:2010} for further improving the phase space sampling and reducing the statistical variances. 

\begin{acknowledgments}
This work was performed using HPC resources from GENCI-[CCRT/CINES] (Grant x2013096973). 
\end{acknowledgments}

\bibliographystyle{prl}

\begin{thebibliography}{}
 \bibitem{chandler:1987} D. Chandler, {\it Introduction to modern statistical mechanics},  Oxford Univ. Press (1987).

 \bibitem{frenkel2002understanding} D. Frenkel and B. Smit, {\it Understanding molecular simulation: from algorithms to applications} Academic Press (2002).  

 \bibitem{wales:2003} D. Wales, {\it Energy Landscapes}, Cambridge University Press, Cambridge (2003).

 \bibitem{jarzynski:1997} C. Jarzynski, Phys. Rev. Lett. {\bf 78}, 2690 (1997).  

 \bibitem{jarzynski:2007} C. Jarzynski, C. R. Physics {\bf 8}, 495 (2007).  

 \bibitem{bennett:1976} C. H. Bennett, J. Comp. Phys.~\textbf{22}, 245 (1976). 

 \bibitem{ferrenberg:1989} A. M. Ferrenberg and R. H. Swendsen, Phys. Rev. Lett.~\textbf{63}, 1195 (1989). 

 \bibitem{tan:2004} Z. Tan, J. Am. Stat. Assoc.~\textbf{99}, 1027 (2004). 

 \bibitem{shirts:2008} M. J. Shirts and J. D. Chodera, J. Chem. Phys.~\textbf{129}, 124105 (2008). 

 \bibitem{habeck:2012} M. Habeck, Phys. Rev. Lett.~\textbf{109}, 100601 (2012). 

 \bibitem{torrie:1977} G. Torrie and J. Valleau, J. Comput. Phys.~\textbf{23}, 187 (1977). 

 \bibitem{kastner:2011} J. K\"{a}stner, Computational Molecular Science~\textbf{932}, (2011). 

 \bibitem{lelievre:2010} T. Leli{\`e}vre, M. Rousset and G. Stoltz, {\it Free-energy computations: a mathematical perspective}, Imperial College Press, 2010.

 \bibitem{carter:1989} E. Carter, G. Ciccotti, J. Hynes and R. Kapral, Chem. Phys. Lett. \textbf{156}, 472 (1989). 

 \bibitem{rosso:2002} L. Rosso, P. Minary, Z. Zhu, M. Tuckerman, J. Chem. Phys.~\textbf{116}, 4389 (2002). 

 \bibitem{darve:2001} E. Darve and A. Pohorille, J. Chem. Phys. \textbf{115}, 9169 (2001). 

 \bibitem{darve:2008} E. Darve, D. Rodriguez-Gomez, A. Pohorille, J. Chem. Phys. \textbf{128}, 144120 (2008).
                                               
 \bibitem{lyubartsev:1992} A. Lyubartsev, A Martinovskii, S. Shevkunov and P. Vorontsov-Velyaminov, J. Chem. Phys. \textbf{96}, 1776 (1992).  

 \bibitem{lelievre:2008} T. Leli\`evre, M. Rousset, G. Stoltz, Nonlinearity \textbf{21}, 1155 (2008). 

% \bibitem{forte:2012} G. Fort, B. Jourdain, E. Kuhn, T. Leli\`evre and G. Stoltz, arXiv:1207.6880.  

 \bibitem{lelievre:2008b} T. Leli\`evre, M. Rousset and G. Stoltz, J. Chem. Phys. \textbf{126}, 134111 (2007). 

 \bibitem{henin:2010} J. H\'enin, J. Fiorin, C. Chipot, M. Klein, J. Chem. Theory Comput. \textbf{6}, 35 (2010). 

 \bibitem{brukhno:1996} A. Brukhno, T. Kuznetsova, A Lyubartsev and P. Vorontsov-Vel'yaminov, Polymer Science A, \textbf{38}, 64 (1996). 

 \bibitem{wang:2001} F. Wang and D. P. Landau, Phys. Rev. Lett. \textbf{86}, 2050 (2001).

 \bibitem{laio:2002} A. Laio and M. Parrinello, Proc. Natl. Acad. Sci. U.S.A. \textbf{99}, 12562 (2002). 

 \bibitem{marsili:2006} S. Marsili, A. Barducci, R. Chelli, P. Porcacci and V. Schettino, J. Phys. Chem. B \textbf{110}, 14011 (2006). 

 \bibitem{lynden-bell:1994} R. M. Lynden-Bell and D. J. Wales, J. Chem. Phys. \textbf{101}, 1460 (1994). 

 \bibitem{lee:1993} J. Lee, Phys. Rev. Lett. \textbf{71}, 211 (1993). 

 \bibitem{ackland2004} G. Ackland, M. Medelev, D. Srolovitz, S. Han and  A. Barashev, J. Phys.: Condens. Matter \textbf{16}, 2629 (2004). 

 \bibitem{athenes:2010} M. Ath\`{e}nes and M.-C. Marinica, J. Comput. Phys. \textbf{229}, 7129 (2010). 
 
 \bibitem{athenes:2012} M. Ath\`{e}nes, M.-C. Marinica and T. Jourdan J. Chem. Phys. \textbf{137}, 194107 (2012). 

 \bibitem{otter:1998} W. den Otter, W. Briels, J. Chem. Phys. \textbf{109}, 4139 (1998). 

 \bibitem{sprik:1998} M. Sprik, G. Ciccotti, J. Chem. Phys. \textbf{109}, 7737 (1998).

 \bibitem{ciccotti:2008} G. Ciccotti, T. Leli\`evre, E. Vanden-Eijnden, Commun. Pure Appl. Math.~\textbf{61}, 371 (2008). 

 \bibitem{frenkel:2004} D. Frenkel, Proc. Natl. Acad. Sci. U.S.A.~{\bf 101}, 17571 (2004).

 \bibitem{frenkel:2006} D. Frenkel, Waste-recycling Monte Carlo, in ``Computer Simulations in Condensed Matter Systems'', Lect. Notes Phys. {\bf 703}, 127 (2006). 

 \bibitem{athenes:2007} M. Ath\`enes, Eur. Phys. J. B \textbf{58}, 83 (2007). 

\end{thebibliography}

\end{document}